\documentclass[oldversion]{aa}  
\usepackage{graphicx,float}
\usepackage{latexsym,amsmath,amssymb}
\usepackage{natbib}
\usepackage{changebar}
\usepackage{subfigure}
\usepackage{txfonts}
\usepackage{rotating}
\usepackage{longtable, lscape}

\usepackage{natbib}
\bibpunct{(}{)}{;}{a}{}{,}

\def\apjl{ApJ}%
\def\apjs{ApJS}%
\def\ao{Appl.~Opt.}%
\def\aap{A\&A}%
\begin{document}

   \title{Luminosity-dependent unification of Active Galactic Nuclei and the X-ray Baldwin effect}
   \subtitle{}

   \author{C. Ricci
          \inst{1,2,3},
          S. Paltani
          \inst{1,2},
          H. Awaki
          \inst{4},
          P.-O. Petrucci
          \inst{5},
          Y. Ueda
          \inst{3},
          and
          M. Brightman
          \inst{6}
          }

   \institute{ \textsl{ISDC} Data Centre for Astrophysics, Universit\'e de Gen\`eve, ch. d'Ecogia 16, 1290 Versoix, Switzerland 
    \and Observatoire de Gen\`eve, Universit\'e de Gen\`eve, 51 Ch. des Maillettes, 1290 Versoix, Switzerland
    \and Department of Astronomy, Kyoto University, Oiwake-cho, Sakyo-ku, Kyoto 606-8502
    \and Department of Physics, Ehime University, Matsuyama, 790-8577, Japan 
    \and UJF-Grenoble 1 / CNRS-INSU, Institut de Plan\'etologie et d'Astrophysique de Grenoble (IPAG) UMR 5274, Grenoble, F-38041, France 
    \and Max-Planck-Institut f\"{u}r extraterrestrische Physik, Giessenbachstrasse 1, D-85748, Garching bei M\"{u}nchen, Germany\\
             }
    \offprints{ricci@kusastro.kyoto-u.ac.jp} 
   \authorrunning{ C. Ricci et al.}
   \titlerunning{Luminosity-dependent unification of AGN and the X-ray Baldwin effect}
    \date{Received; accepted}

 
 \abstract{The existence of an anti-correlation between the equivalent width ($EW$) of the narrow core of the iron K$\alpha$ line and the luminosity of the continuum (i.e. the X-ray Baldwin effect) in type-I active galactic nuclei has been confirmed over the last years by several studies carried out with {\it XMM-Newton}, {\it Chandra} and {\it Suzaku}. However, so far no general consensus on the origin of this trend has been reached. Several works have proposed the decrease of the covering factor of the molecular torus with the luminosity (in the framework of the luminosity-dependent unification models) as a possible explanation for the X-ray Baldwin effect. Using the fraction of obscured sources measured by recent X-ray and IR surveys as a proxy of the half-opening angle of the torus, and the recent Monte-Carlo simulations of the X-ray radiation reprocessed by a structure with a spherical-toroidal geometry by Ikeda et al. (2009) and Brightman \& Nandra (2011), we test the hypothesis that the X-ray Baldwin effect is related to the decrease of the half-opening angle of the torus with the luminosity. Simulating the spectra of an unabsorbed population with a luminosity-dependent covering factor of the torus as predicted by recent X-ray surveys, we find that this mechanism is able to explain the observed X-ray Baldwin effect. Fitting the simulated data with a log-linear $L_{\,\mathrm{2-10\,keV}}-EW$ relation, we found that in the Seyfert regime ($L_{\,2-10\rm\,keV}\leq 10^{44.2}\rm\,erg\,s^{-1}$) luminosity-dependent unification produces a slope consistent with the observations for average values of the equatorial column densities of the torus of $\log N_{\rm\,H}^{\rm\,T}\gtrsim 23.1$, and can reproduce both the slope and the intercept for $ \log N_{\rm\,H}^{\rm\,T}\simeq 23.2$. Lower values of $N_{\rm\,H}^{\rm\,T}$ are obtained considering the decrease of the covering factor of the torus with the luminosity extrapolated from IR observations ($22.9 \lesssim \log N_{\rm\,H}^{\rm\,T}\lesssim 23$). In the quasar regime ($L_{\,2-10\rm\,keV}> 10^{44.2}\rm\,erg\,s^{-1}$) a decrease of the covering factor of the torus with the luminosity slower than that observed in the Seyfert regime (as found by recent hard X-ray surveys) is able to reproduce the observations for $23.2\lesssim \log N_{\rm\,H}^{\rm\,T}\lesssim 24.2$.

 }

   \maketitle
   
\section{Introduction}

The observational signatures of reprocessed radiation in the X-ray spectra of Active Galactic Nuclei (AGN) are mainly two: a Compton hump peaking around 30\,keV and a fluorescent iron K$\alpha$ line. While the Compton hump is produced only if the reprocessing material is Compton-thick (CT, $N_{\rm\,H}\gtrsim 10^{24}\rm\,cm^{-2}$), the iron K$\alpha$ line is created also in Compton-thin material (e.g., \citealp{Matt:2003fk}). The iron K$\alpha$ line is often observed as the superposition of two different components: a broad and a narrow one. The broad component has a full width at half-maximum (FWHM) of $\gtrsim 30,000\rm\,km\,s^{-1}$, and is thought to be created close to the black hole in the accretion disk (e.g., \citealp{Fabian:2000uq}), or to be related to the presence of features created by partially covering warm absorbers in the line of sight (e.g., \citealp{Turner:2009kx}, \citealp{Miyakawa:2012fk}). While the broad component is observed in only $\sim 35-45\%$ of bright nearby AGN (e.g., \citealp{de-La-Calle-Perez:2010fk}), the narrow ($\mathrm{FWHM}\sim 2,000\rm\,km\,s^{-1}$, \citealp{Shu:2010zr}) core of the iron line has been found to be almost ubiquitous (e.g., \citealp{Nandra:2007ly}, \citealp{Singh:2011ly}). This component peaks at $6.4$\,keV (e.g.,\citealp{Yaqoob:2004vn}), which points to the line being produced in cold neutral material. This material has often been identified as circumnuclear matter located at several thousand gravitational radii from the supermassive black hole, and is likely related to the putative molecular torus (e.g., \citealp{Nandra:2006fk}), although a contribution of the outer part of the disk (e.g., \citealp{Petrucci:2002fk}) or of the broad-line region (BLR, e.g., \citealp{Bianchi:2008fk}) cannot be excluded. In a recent study, \citet{Shu:2011fk} found that the weighted mean of the ratio between the FWHM of the narrow Fe K$\alpha$ line and that of optical lines produced in the BLR is $\simeq 0.6$. This implies that the size of the iron K$\alpha$-emitting region is on average $\sim3$ times that of the BLR, and points towards most of the narrow iron K$\alpha$ emission being produced in the putative molecular torus.

\begin{table*}
\caption[]{Summary of the most recent studies (along with the original work of \citealp{Iwasawa:1993ys}) of the X-ray Baldwin effect. The model commonly used to fit the data is $\log EW=A+B \log L_{\mathrm{X,44}}$, where $L_{\mathrm{X,44}}$ is the 2--10\,keV luminosity in units of $10^{44}\rm\,erg\,s^{-1}$. The table is divided in works for which the fit was done {\it per observation} (i.e. using all the available observations for every source of the sample) and {\it per source} (i.e. averaging the values of $EW$ and $L_{\,\mathrm{X}}$ of different observations of the same source). The table lists the (1) reference, (2) the value of the intercept and (3) of the slope, (4) the number and the type of objects in the sample, and (5) the observatory and instrument used. Dashes are reported when the value is not available.}
\label{tab:XBEref}
\begin{center}
\begin{tabular}{lcccc}
\hline
\hline
\noalign{\smallskip}
\multicolumn{1}{c}{(1)} & (2) & (3) & (4) & (5)\\ 
\noalign{\smallskip}
Reference & $A$ & $B$ & Sample & Observatory/instrument\\ 
\noalign{\smallskip}
\hline
\noalign{\smallskip}
\multicolumn{5}{l}{\bf Fits per Observation} \\
\noalign{\smallskip}
\hline
\noalign{\smallskip}
\noalign{\smallskip}
\citealp{Shu:2012fk}       &   $1.58\pm0.03^*$  & $-0.18\pm0.03$ & 32 (RQ) & {\it Chandra}/HEG \\  
\noalign{\smallskip}
\citealp{Shu:2010zr}       &   $1.58\pm 0.03^*$ & $-0.22\pm0.03$ & 33 (RQ+RL) & {\it Chandra}/HEG \\  
\noalign{\smallskip}
  \citealp{Bianchi:2007vn}     &  $1.73\pm 0.03$ & $-0.17\pm0.03$ & 157 (RQ) & {\it XMM-Newton}/EPIC \\                    		
\noalign{\smallskip}
 \citealp{Jiang:2006vn}  &  -- & $-0.20\pm0.04$ & 101 (RL+RQ)& {\it XMM-Newton}/EPIC + {\it Chandra}/HEG \\  
\noalign{\smallskip}
 \citealp{Jiang:2006vn}  & --  & $-0.10\pm0.05$ & 75 (RQ) & {\it XMM-Newton}/EPIC + {\it Chandra}/HEG \\  
\noalign{\smallskip}
   \citealp{Jimenez-Bailon:2005tg}    &  -- & $-0.06\pm0.20$ & 38 (RQ) &  {\it XMM-Newton}/EPIC \\                    		
 \noalign{\smallskip}
 \citealp{Zhou:2005ys}  &  -- &$-0.19\pm0.04$ & 66 (RQ+RL)& {\it XMM-Newton}/EPIC \\ 
 \noalign{\smallskip}
     \citealp{Page:2004kx}    &  -- &$-0.17\pm0.08$ & 53 (RQ+RL)& {\it XMM-Newton}/EPIC \\                    		
\noalign{\smallskip}
\citealp{Iwasawa:1993ys} &  --  &$-0.20\pm0.03$ & 37 (RQ+RL)& {\it Ginga} \\      
\noalign{\smallskip}

\noalign{\smallskip}
\hline
\noalign{\smallskip}
\multicolumn{5}{l}{\bf Fits per Source} \\
\noalign{\smallskip}
\hline
\noalign{\smallskip}
\citealp{Shu:2012fk}       &   $1.64\pm0.03^*$  & $-0.11\pm0.03$ & 32 (RQ) & {\it Chandra}/HEG \\  
\noalign{\smallskip}
\citealp{Shu:2010zr}       &  $1.63\pm 0.04^*$ & $-0.13\pm0.04$ & 33 (RQ+RL) & {\it Chandra}/HEG \\                    
\noalign{\smallskip}
\hline
\noalign{\smallskip}
\multicolumn{5}{l}{{\tiny {\bf Notes}: $^*$ These values of the intercepts were corrected in order to be consistent with Eq.\,\ref{eq:fit_XBE}, as those reported in \citet{Shu:2010zr,Shu:2012fk}  } } \\
\multicolumn{5}{l}{{\tiny  are relative to the 2--10\,keV luminosity in units of $10^{43}\rm\,erg\,s^{-1}$ ($L_{X,43}$). }} \\
\end{tabular}
\end{center}
\end{table*}

One of the most interesting characteristics of the narrow component of the iron K$\alpha$ line is the inverse correlation between its equivalent width ($EW$) and the X-ray luminosity (e.g., \citealp{Iwasawa:1993ys}). The existence of an anti-correlation between the equivalent width of a line and the luminosity of the AGN continuum was found for the first time in the UV by \citet{Baldwin:1977fk} for the C\,IV$\,\lambda 1549$ line, and was then dubbed the {\it Baldwin effect}. A similar trend was later found for several other emission lines such as Ly$\alpha$, [C\,III]$\,\lambda 1908$, Si\,IV$\,\lambda 1396$, Mg\,II\,$\lambda 2798$ \citep{Dietrich:2002fk}, UV iron emission lines \citep{Green:2001uq}, mid-IR lines such as [AR\,III]$\,\lambda8.99\mu m$, [S\,IV]$\,\lambda10.51\mu m$ and [Ne\,II]$\,\lambda12.81\mu m$ \citep{Honig:2008kx}, and forbidden lines as [O\,II]$\,\lambda 3727$ and [Ne\,V]$\,\lambda 3426$ \citep{Croom:2002vn}. The slope of the Baldwin effect for most of these lines has been shown to be steeper for those originating from higher ionization species (e.g., \citealp{Dietrich:2002fk}). The origin of the Baldwin effect is still unknown, although several possible explanations have been put forward, and it might be different for lines originating in different regions of the AGN. For the lines produced in the BLR it might be related to the lower ionization and photoelectric heating in the BLR gas of more luminous objects (e.g., \citealp{Netzer:1992fk}).

In the X-rays, \citet{Iwasawa:1993ys} using {\it Ginga} observations of 37 AGN, found the first evidence of an anti-correlation between the equivalent width of the iron K$\alpha$ line and the $2-10\rm\,keV$ luminosity ($EW \propto L_{\,\mathrm{X}}^{-0.20\pm0.04}$). This trend is usually called the {\it X-ray Baldwin effect} or the {\it Iwasawa-Taniguchi effect}.
Using {\it ASCA} observations, \citet{Nandra:1997ve} confirmed the existence of such an anti-correlation, and argued that most of the effect could be explained by variations of the broad component of the iron K$\alpha$ line with the luminosity. The advent of {\it XMM-Newton}, {\it Chandra} and {\it Suzaku} made however clear that most of the observed X-ray Baldwin effect is due to the narrow core of the iron K$\alpha$ line (e.g., \citealp{Page:2004kx}, \citealp{Shu:2010zr}, \citealp{Fukazawa:2011ly}).
The significance of the effect was questioned by \citet{Jimenez-Bailon:2005tg}, who discussed the possible importance of contamination from radio-loud (RL) AGN, which have on average larger X-ray luminosities and smaller signatures of reprocessed radiation (e.g., \citealp{Reeves:2000uq}).
However, \citet{Grandi:2006zr}, studying {\it BeppoSAX} observations of radio-loud AGN, also found evidence of an X-ray Baldwin effect. This, together with the study of a large sample of radio-quiet (RQ) AGN carried out by \citet{Bianchi:2007vn} using {\it XMM-Newton} data, showed that the X-ray Baldwin effect is not a mere artifact. \citet{Jiang:2006vn} suggested that the X-ray Baldwin effect might be due to the delay between the variability of the AGN primary energy source and that of the reprocessing material located farther away. Using {\it Chandra}/HEG observations, \citet{Shu:2010zr,Shu:2012fk} confirmed the importance of variability, showing that averaging the values of $EW$ obtained over multiple observations of individual sources would significantly attenuate the anti-correlation from $EW \propto L_{\,\mathrm{X}}^{-0.18\pm0.03}$ to $EW \propto L_{\,\mathrm{X}}^{-0.11\pm0.03}$. {\it Chandra}/HEG observations \citep{Shu:2010zr} have also shown that the normalization of the X-ray Baldwin effect is lower (i.e. the average $EW$ of the narrow Fe K$\alpha$ component is smaller) with respect to that obtained by previous works performed using {\it XMM-Newton}. In Table\,\ref{tab:XBEref} we report the values of the slope and the intercept obtained from the most recent {\it XMM-Newton}/EPIC and {\it Chandra}/HEG works, along with the original {\it Ginga} work of \citet{Iwasawa:1993ys}.

\begin{figure*}[t!]
\centering
\begin{minipage}[!b]{.48\textwidth}
\centering
\includegraphics[angle=270,width=9cm]{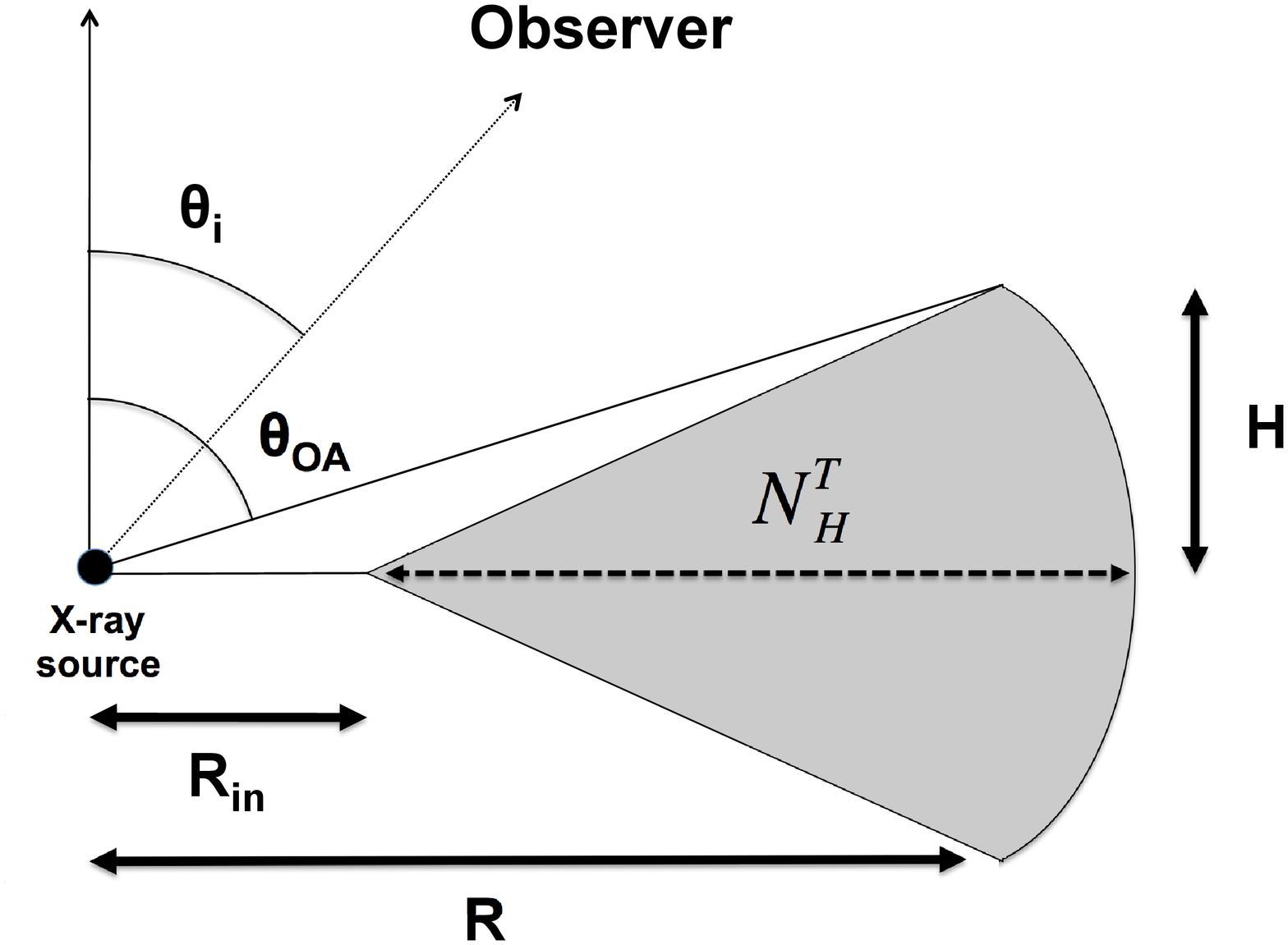}
 \caption{Schematic representation of the torus geometry considered. The angle $\theta_{\,\mathrm{i}}$ is the inclination of the line of sight with respect to the torus axis, while $\theta_{\mathrm{OA}}$ and $N_{\rm\,H}^{\rm\,T}$ are the half-opening angle and the equatorial column density of the molecular torus, respectively. $H$ is the maximum height of the torus, while $R_{\mathrm{in}}$ and $R$ are its inner and outer radius, respectively.}\label{fig:tor_CFL1}
\end{minipage}
\hspace{0.05cm}
\begin{minipage}[!b]{.48\textwidth}
\centering
\includegraphics[width=9cm]{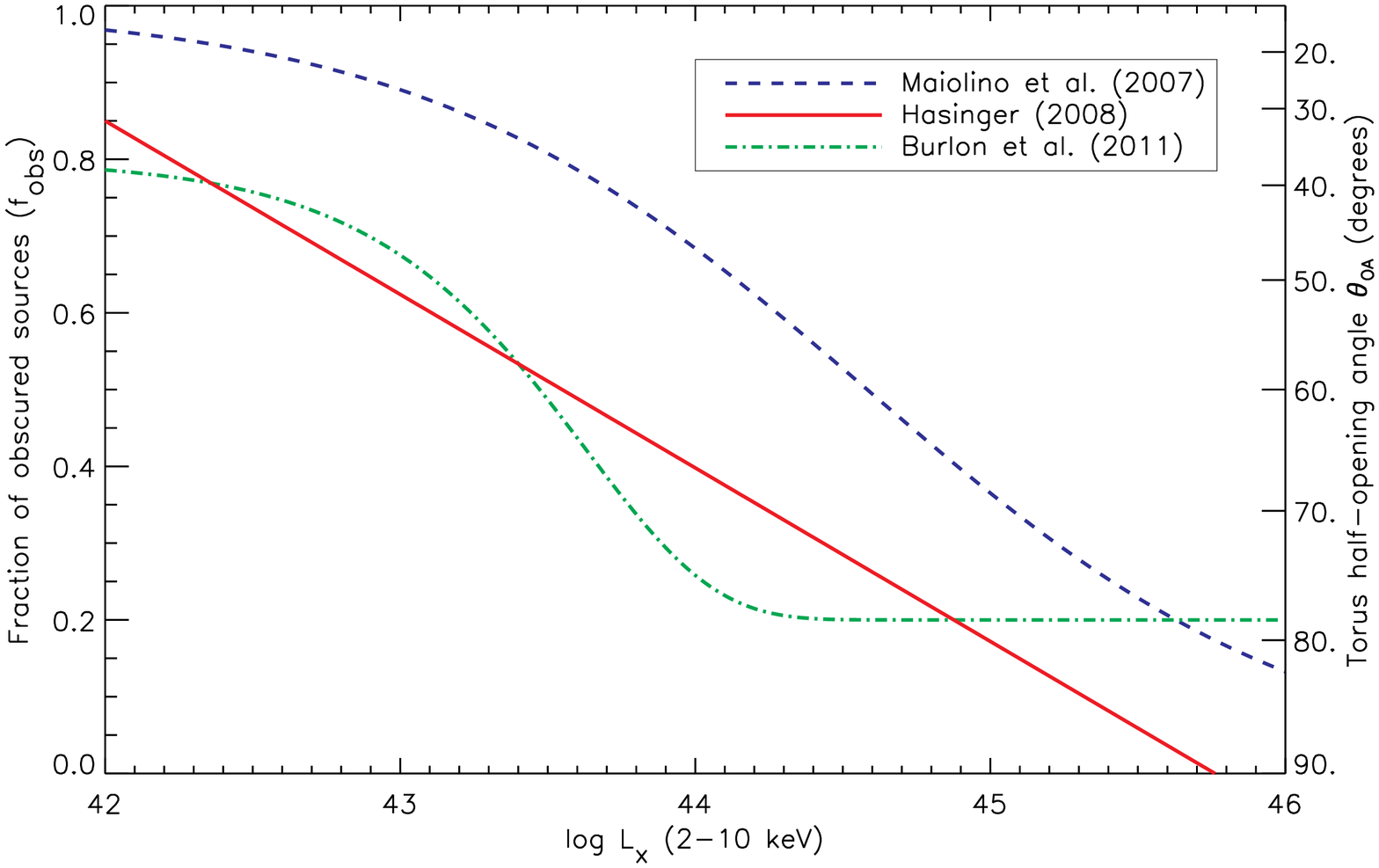}
 \caption{Variation of the fraction of obscured sources ($f_{\mathrm{obs}}$) and of the half-opening angle of the torus ($\theta_{\mathrm{OA}}$) with the 2--10\,keV luminosity for three of the most recent medium X-ray (2--10\,keV, \citealp{Hasinger:2008ve}), hard X-ray (15--55\,keV, \citealp{Burlon:2011cr}) and IR \citep{Maiolino:2007bh} studies (Eq.\,\ref{Eq:Hasinger}-\ref{Eq:Maiolino}). }\label{fig:tor_CFL2}\end{minipage}

\end{figure*}

An intriguing possibility is that the X-ray Baldwin effect is related to the decrease of the covering factor of the torus with luminosity (e.g., \citealp{Page:2004kx}, \citealp{Zhou:2005ys}), in the frame of the {\it luminosity-dependent unification} schemes. The equivalent width of the iron K$\alpha$ line is in fact expected to be proportional to the covering factor of the torus (e.g., \citealp{Krolik:1994fk}, \citealp{Ikeda:2009nx}), so that a covering factor of the torus decreasing with the luminosity might in principle be able to explain the X-ray Baldwin effect. Luminosity-dependent unification models have been proposed to explain the decrease of the fraction of obscured objects ($f_{\mathrm{obs}}$) with the increase of the AGN output power. The first suggestion of the existence of a relation between $f_{\mathrm{obs}}$ and the luminosity came about 30 years ago \citep{Lawrence:1982ys}. Since then the idea that the covering factor of the obscuring material decreases with luminosity has been gaining observational evidence from radio (e.g., \citealp{Grimes:2004kx}), infrared  (e.g., \citealp{Treister:2008uq}, \citealp{Mor:2009fk}, \citealp{Gandhi:2009uq}), optical (e.g., \citealp{Simpson:2005uq}) and X-ray (e.g., \citealp{Ueda:2003qf}, \citealp{Beckmann:2009fk}, \citealp{Ueda:2011fk}) studies of AGN. Although luminosity-dependent unification models have been suspected for long to play a major role in the X-ray Baldwin effect, so far no quantitative estimation of this effect has been performed.

In this work we quantify for the first time the influence of the decrease of the covering factor of the torus with the luminosity on the equivalent width of the iron K$\alpha$ line. Using the Monte-Carlo spectral simulations of a torus with a variable half-opening angle ($\theta_{\mathrm{OA}}$) recently presented by \citet{Ikeda:2009nx} and \citet{Brightman:2011oq}, together with the most recent and comprehensive observations in different energy bands of the decrease of $f_{\mathrm{obs}}$ with the luminosity, we show that this mechanism is able to explain the X-ray Baldwin effect. The paper is organized as follows. In Sect.\,\ref{Sec:Simulations} we present in detail our spectral simulations, and in Sect.\,\ref{sect:XBEslope} we compare the slopes and intercepts obtained by our simulations with those measured by high sensitivity {\it Chandra}/HEG observations, as they provide the best energy resolution available to date. In Sect.\,\ref{Sec:discussion} we discuss our results, and in Sect.\,\ref{Sec:summary} we present our conclusions.

\section{Simulations}\label{Sec:Simulations}

\subsection{The relation between $\theta_{\mathrm{OA}}$ and $L_{\mathrm{X}}$}

As a proxy of the relationship between the half-opening angle of the molecular torus $\theta_{\mathrm{OA}}$ (see Fig.\,\ref{fig:tor_CFL1}) and the intrinsic X-ray luminosity of AGN we used the variation of the fraction of obscured sources $f_{\mathrm{obs}}$ with luminosity measured by the recent medium (2--10\,keV) and hard (15--55\,keV) X-ray surveys of \citet{Hasinger:2008ve} and \citet{Burlon:2011cr}.
In the 2--10\,keV band, from a combination of surveys performed by {\it HEAO-1}, {\it ASCA}, {\it BeppoSAX}, {\it XMM-Newton} and {\it Chandra} in the luminosity range $42\leq \log L_{\mathrm{X}} \leq 46$, \citet{Hasinger:2008ve} found 
\begin{equation}\label{Eq:Hasinger}
f_{\mathrm{obs}}\simeq-0.226\log L_{\mathrm{X}}+10.342, 
\end{equation}
where $L_{\mathrm{X}}$ is the luminosity in the 2--10\,keV energy range. In the 15--55\,keV band, using {\it Swift}/BAT to study AGN in the luminosity range $42\leq \log L_{\mathrm{X}} \leq 45$, and fitting the data with
\begin{equation}\label{Eq:Burlon} 
f_{\mathrm{obs}}\simeq R_{\mathrm{low}}\,e^{-L_{\mathrm{HX}}/L_{\mathrm{C}}}+R_{\mathrm{high}}\,(1-e^{-L_{\mathrm{HX}}/L_{\mathrm{C}}}),
\end{equation}
where $L_{\mathrm{HX}}$ is the luminosity in the 15--55\,keV band, \citet{Burlon:2011cr} obtained $R_{\mathrm{low}}=0.8$, $R_{\mathrm{high}}=0.2$ and $L_{\mathrm{C}}=10^{43.7}\rm\,erg\,s^{-1}$. We also used the results of the IR work of \citet{Maiolino:2007bh}, who found an anti-correlation between the ratio $\lambda L_{\lambda}(6.7\mu m)/\lambda L_{\lambda}(5100\AA)$ and the [O\,III]$\lambda 5007\AA$ line luminosity. This trend was interpreted as an effect of the decrease of the covering factor of the circumnuclear dust as a function of luminosity, with $f_{\mathrm{obs}}$ varying as
 \begin{equation}\label{Eq:Maiolino} 
 f_{\mathrm{obs}}\simeq \frac{1}{1+L_{\mathrm{opt}}^{0.414}},
 \end{equation} 
where $L_{\mathrm{opt}}=(\lambda L_{\lambda}(5100\AA))/10^{45.63}$. Following \citet{Maiolino:2007bh} it is possible to rewrite $L_{\mathrm{opt}}$ as a function of the X-ray luminosity: $L_{\mathrm{opt}}=L_{\mathrm{X}}^{1.39}/10^{61.97}$.

The fraction of obscured sources at a given luminosity can be easily related to the half-opening angle of the torus using 
\begin{equation}\label{Eq:fractioncos}
\theta_{\mathrm{OA}}=\cos ^{-1}( f_{\mathrm{obs}}).
\end{equation} 
For Eq.\,\ref{Eq:Burlon} we converted the hard X-ray luminosity to the luminosity in the 2--10\,keV band assuming a photon index $\Gamma=1.9$ (e.g., \citealp{Beckmann:2009fk}). The three $\theta_{\mathrm{OA}}-L_{\mathrm{X}}$ relationships used in this work are shown in Fig.\,\ref{fig:tor_CFL2}.

\begin{figure}[t!]
\centering
\includegraphics[width=8.5cm]{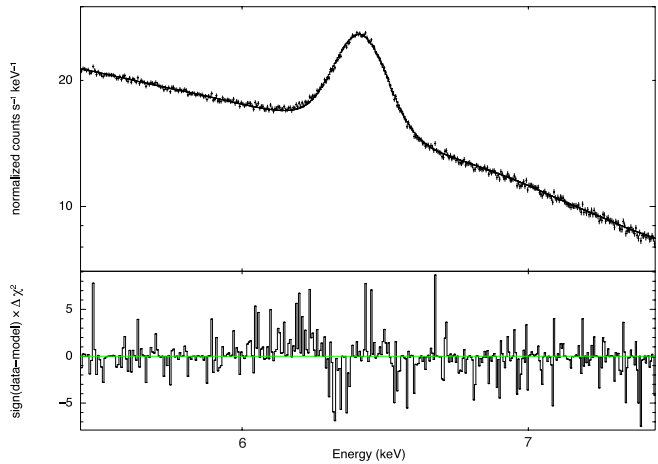}
\caption{{\it Top panel:} extract in the 5.5-7.5\,keV region of a spectrum simulated using the model of \citet{Ikeda:2009nx}. The model has a torus with an half-opening angle of $\theta_{\mathrm{OA}}=46.2^{\circ}$ (equivalent to $\log L_{\mathrm{X}}=42.7$ according to the relationship of \citealp{Hasinger:2008ve}), an equatorial column density of $\log N_{\rm\,H}^{\rm\,T}=23.8$, and an inclination angle of $\theta_{\,\mathrm{i}}=1^{\circ}$. The continuous line represents the fit to the simulated spectrum using for the continuum the same model we used for the simulations, and a Gaussian line for the iron K$\alpha$ fluorescent line. {\it Bottom panel:} contribution to the chi-squared for the best fit to the simulations. }
\label{fig:simulations_line}
\end{figure}%

\subsection{Spectral simulations and fitting}

\citet{Ikeda:2009nx} recently presented Monte-Carlo simulations of the reprocessed X-ray emission of an AGN surrounded by a three-dimensional spherical-toroidal structure. The simulations were performed using the ray-tracing method, taking into account Compton down-scattering and absorption, and are stored in tables, so that they can be used to perform spectral fitting. The free parameters of this model are the half-opening angle of the torus $\theta_{\mathrm{OA}}$, the line of sight inclination angle $\theta_{\,\mathrm{i}}$, the torus equatorial column density $N_{\rm\,H}^{\rm\,T}$ and the photon index $\Gamma$ of the primary continuum. {\it Note that $N_{\rm\,H}^{\rm\,T}$ should not be confused with the observed hydrogen column density $N_{\rm\,H}$. In all objects, $N_{\rm\,H}\leq N_{\rm\,H}^{\rm\,T}$, with $N_{\rm\,H}=N_{\rm\,H}^{\rm\,T}$ only if $\theta_{\,\mathrm{i}}=90^{\circ}$}. In Fig.\,\ref{fig:tor_CFL1} a schematic representation of the geometry considered is shown. The dependence of $N_{\rm\,H}$ on the inclination angle is given by Eq.\,3 of \citet{Ikeda:2009nx}. In the model of \citet{Ikeda:2009nx} the ratio $R_{\mathrm{in}}/R$ is fixed to 0.01, and the inclination angle of the observer ${\mathit \theta}_{\mathrm{i}}$ can vary between $1^{\circ}$ and $89^{\circ}$, $N_{\rm\,H}^{\rm\,T}$ between $10^{22}$ and $10^{25}\rm\,cm^{-2}$, while $\theta_{\mathrm{OA}}$ spans the range between $10^{\circ}$ and $70^{\circ}$. Because of the assumed dependence of the half-opening angle on the luminosity, the interval of values of $\theta_{\mathrm{OA}}$ allowed by the model limits the range of luminosities we can probe. In the following we will consider luminosities above $\log L_{\mathrm{X}} = 42$, as below this value few AGN are detected, and several works point towards a possible disappearance of the molecular torus (e.g., \citealp{Elitzur:2006fk}). The upper-limit luminosity we can reach depends on the $\theta_{\mathrm{OA}}-L_{\mathrm{X}}$ relationship used, and considering $\theta_{\mathrm{OA}}^{\mathrm{\,max}}=70^{\circ}$, is $\log L_{\rm\,max}=44.2$ for \citet{Hasinger:2008ve}, $\log L_{\rm\,max}=43.8$ for \citet{Burlon:2011cr} and $\log L_{\rm\,max}=45.0$ for \citet{Maiolino:2007bh}. Due to its limitations in the values of $\theta_{\mathrm{OA}}$ permitted, the model of \citet{Ikeda:2009nx} restricts most of the simulations to the Seyfert regime ($\log L_{\mathrm{X}}\leq \log L_{\mathrm{X}}^{\mathrm{Q}}=44.2$). 

To extend our study to the quasar regime ($L_{\mathrm{X}}> L_{\mathrm{X}}^{\mathrm{Q}}$) we simulated the X-ray Baldwin effect using the spectral model of \citet{Brightman:2011oq}. This model considers the same geometry as \citet{Ikeda:2009nx}, but has $R_{\mathrm{in}}=0$, a $\theta_{\,\mathrm{i}}$-independent $N_{\rm\,H}$, and it allows different values of $\theta_{\mathrm{OA}}$ ($26^{\circ}$ to $84^{\circ}$). The higher maximum half-opening angle permitted by this model allows to reach higher maximum luminosities in the simulations for the relationships of \citeauthor{Hasinger:2008ve} (\citeyear{Hasinger:2008ve}; $\log L_{\rm\,max}=45.3$) and \citeauthor{Maiolino:2007bh} (\citeyear{Maiolino:2007bh}; $\log L_{\rm\,max}=46.2$), while that of \citet{Burlon:2011cr} is flat ($f_{\rm\,obs}\simeq0.2$) in the quasar regime. However, the lower boundary of $\theta_{\mathrm{OA}}$ in the model of \citet{Brightman:2011oq} limits the lower luminosity we can reach in the simulations, in particular $\log L_{\rm\,min}\simeq 43$ for the relation of \citet{Maiolino:2007bh}. Thus, we used the model of \citet{Ikeda:2009nx} to study the X-ray Baldwin effect in the Seyfert regime ($42\leq\log L_{\mathrm{X}} \leq 44.2$), and the model of \citet{Brightman:2011oq} to probe the quasar regime ($\log L_{\mathrm{X}} > 44.2$).

\begin{figure}[t!]
\centering
\includegraphics[width=9cm]{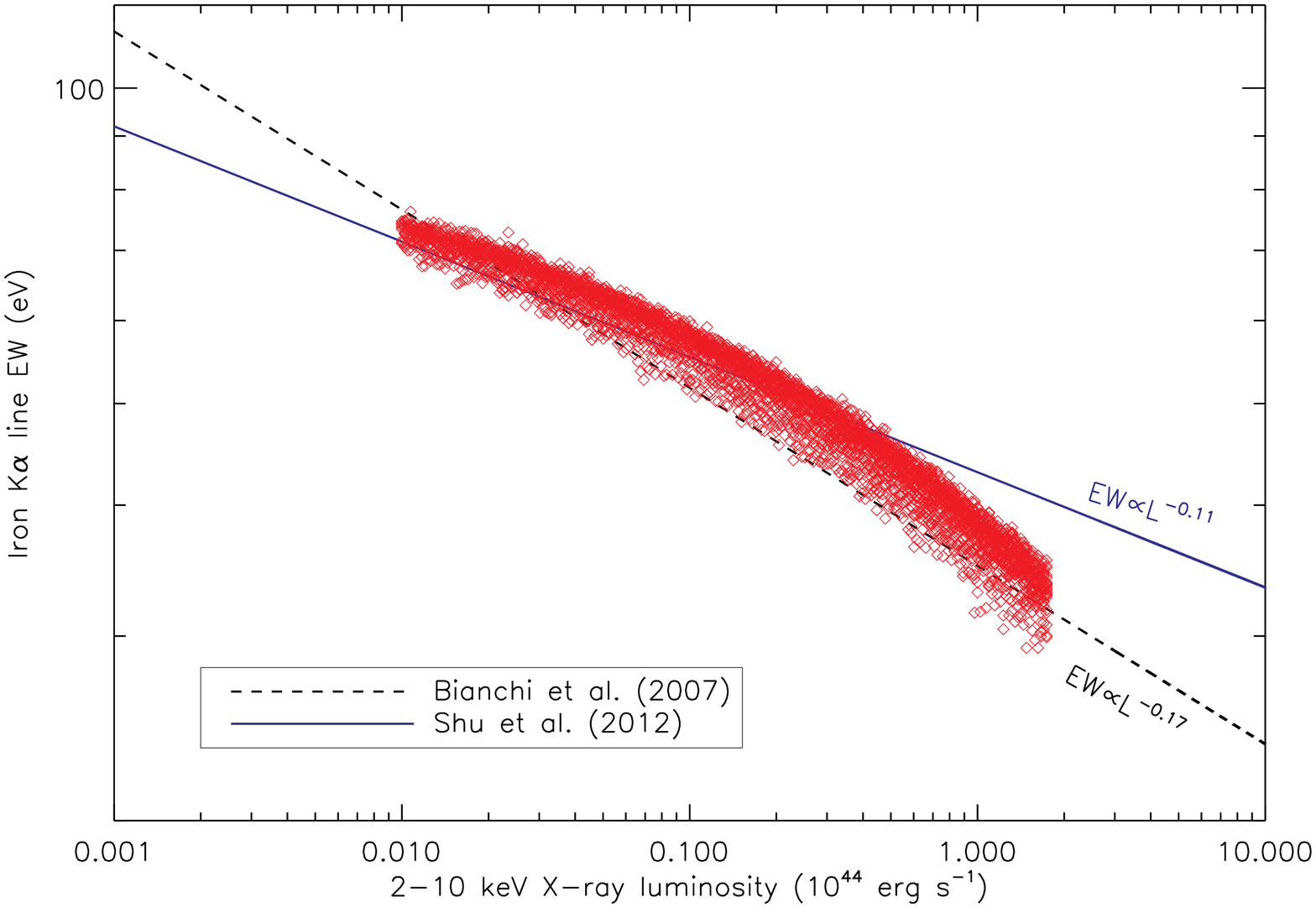}
\caption{Equivalent width of the iron K$\alpha$ line versus the X-ray luminosity obtained simulating a torus with an equatorial column density of $\log N_{\rm\,H}^{\rm\,T}=23.1$ and a covering factor decreasing with the luminosity. The $\theta_{\mathrm{OA}}-L_{\mathrm{X}}$ relationship used here is that of \citet{Hasinger:2008ve} (see Fig.\,\ref{fig:tor_CFL2}). The relations of \citet{Bianchi:2007vn} (black dashed line) and \citeauthor{Shu:2012fk} (\citeyear{Shu:2012fk}; blue line, obtained averaging different observations of the same source) are also shown for comparison (see also Table\,\ref{tab:XBEref}). The normalization of the relation of \citet{Bianchi:2007vn} has been fixed to an arbitrary value for comparison.}
\label{fig:IT_hasinger}
\end{figure}%

\begin{figure*}[t!]
\centering
\begin{minipage}[!b]{.48\textwidth}
\centering
\includegraphics[width=9cm]{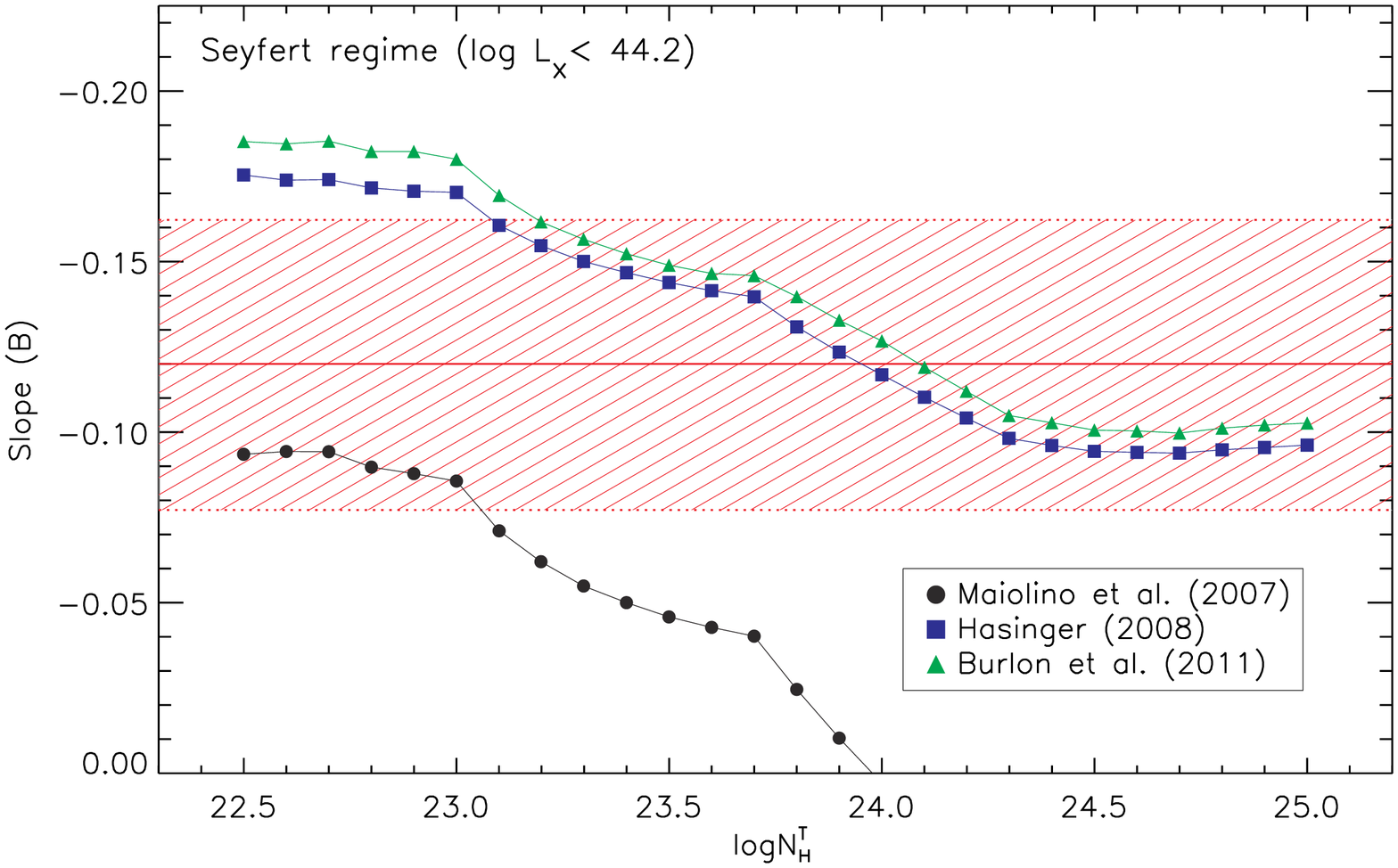}
\end{minipage}
\hspace{0.05cm}
\begin{minipage}[!b]{.48\textwidth}
\centering
\includegraphics[width=9cm]{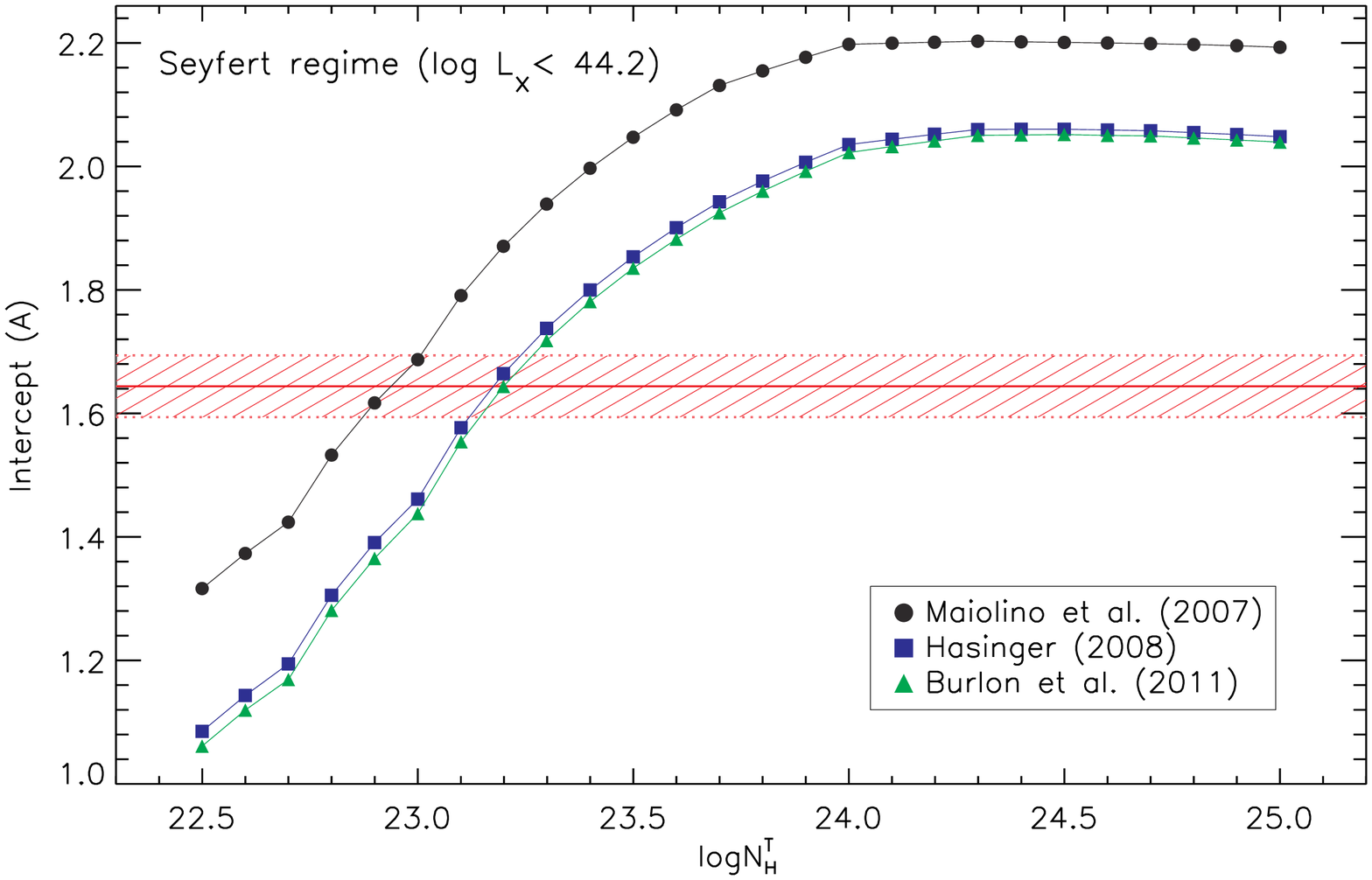}\end{minipage}

 \begin{minipage}[t]{1\textwidth}
  \caption{{\it Left panel}: value of the slope ($B$) of the X-ray Baldwin effect obtained simulating the variation of the reprocessed X-ray radiation with the luminosity for tori with different values of the equatorial column density $N_{\rm\,H}^{\rm\,T}$ (in the range $\log N_{\rm\,H}^{\rm\,T}=22.5-25$), for the different $\theta_{\mathrm{OA}}-L_{\mathrm{X}}$ relationships shown in Fig.\,\ref{fig:tor_CFL2} (Eq.\,\ref{Eq:Hasinger}-\ref{Eq:Maiolino}) in the Seyfert regime ($42 <\log L_{\mathrm{X}} \leq 44.2$). The model used is that of \citet{Ikeda:2009nx}, and the simulated data were fitted with Eq.\,\ref{eq:fit_XBE}. The red line represents the value of the slope ($B=-0.12\pm0.04$) obtained by fitting in the Seyfert regime the data {\it per source} reported in \citet{Shu:2010zr}; the red shadowed area represents the $1\sigma$ contour of the slope. {\it Right panel}: same as left panel, but considering the intercept ($A$) of the X-ray Baldwin effect obtained by our simulations. The red line and shadow represent the value of the intercept obtained by the fit to the {\it Chandra}/HEG data in the Seyfert regime and its $1\sigma$ error ($A=1.64\pm0.05$), respectively.}
\label{fig:slope_NH}
 \end{minipage}

\end{figure*}

To estimate the influence of the decreasing covering factor of the torus on the equivalent width of the iron line, we simulated, using the physical torus models, a large number of spectra using the $\theta_{\mathrm{OA}}-L_{\mathrm{X}}$ relationships reported in Eq.\,\ref{Eq:Hasinger}-\ref{Eq:Maiolino}. Using the three different $\theta_{\mathrm{OA}} - L_{\mathrm{X}}$ relationships, we extrapolated the value of $\theta_{\mathrm{OA}}$ for each luminosity bin ($\Delta \log L_{\mathrm{X}}=0.01$). We fixed the equatorial column density of the torus $N_{\rm\,H}^{\rm\,T}$ to 26 different values, spanning between $10^{22.5}\rm\,cm^{-2}$ and $10^{25}\rm\,cm^{-2}$, with a step of $\Delta \log N_{\rm\,H}^{\rm\,T}=0.1$. To simulate an unabsorbed population, similarly to what is usually used to determine the X-ray Baldwin effect, for each value of $L_{\rm\,X}$ and $N_{\rm\,H}^{\rm\,T}$ we considered inclination angles $\theta_{\,\mathrm{i}}$ between $1^{\circ}$ and $\theta_{\mathrm{OA}}$, with a binning of $\Delta \theta_{\,\mathrm{i}}=3^{\circ}$. In XSPEC 12.7.1 \citep{Arnaud:1996kx} we simulated spectra for each of the three $\theta_{\mathrm{OA}}-L_{\mathrm{X}}$ relationships and for each bin of column density. For the continuum we used a power law with $\Gamma=1.9$ (e.g., \citealp{Beckmann:2009fk}). Our choice of the photon index does not affect significantly the simulations, and adding a scatter of $\Delta \Gamma=0.3$ we found $EW-L_{\mathrm{X}}$ trends consistent with those obtained without scatter. The metallicity was set to the solar value in all our simulations, and the value of the normalization of the reflected component was fixed to that of the continuum.

We fitted the simulated spectra in the 0.3-10\,keV band using the same model adopted for the simulations, substituting the iron K$\alpha$ line component with a Gaussian line. We obtained a good fit for all the simulations, with a reduced chi-squared of $\chi^{2}_{\nu}\lesssim 1.1$. We report in Fig.\,\ref{fig:simulations_line} an example of a typical fit. The model used for the continuum fitting does not affect significantly the results, and using an alternative model like \texttt{pexrav} \citep{Magdziarz:1995pi} the equivalent width of the iron K$\alpha$ line differs on average of only $\sim4\%$. We evaluated the equivalent width of the iron K$\alpha$ line and studied its relationship with the luminosity. As an example, in Fig.\,\ref{fig:IT_hasinger} the simulated X-ray Baldwin effect obtained using the model of \citet{Ikeda:2009nx} and the $\theta_{\mathrm{OA}}-L_{\mathrm{X}}$ relationship of \citet{Hasinger:2008ve} for $\log N_{\rm\,H}^{\rm\,T}=23.1$ is shown, together with the fit to the X-ray Baldwin effect obtained by the recent works of \citet{Bianchi:2007vn} and \citet{Shu:2012fk}. The spread in $EW$ for a given luminosity is due to the range of values of $\theta_{\,\mathrm{i}}$ we considered: larger values of $EW$ usually correspond to lower values of $\theta_{\,\mathrm{i}}$.

Most of the studies of the X-ray Baldwin effect performed in the last years have used a relationship of the type 
\begin{equation}\label{eq:fit_XBE}
\log EW=A+B \log L_{\mathrm{X,44}},
\end{equation}
to fit the $EW-L_{\mathrm{X}}$ trend, where $L_{\mathrm{X,44}}$ is the luminosity in units of $10^{44}\rm\,erg\,s^{-1}$. In order to compare the simulated the $EW-L_{\mathrm{X}}$ trend with that observed in unabsorbed populations of AGN, we fitted, for each value of $N_{\rm\,H}^{\rm\,T}$, the simulated data with Eq.\,\ref{eq:fit_XBE} using the weighted least-square method, with weights of $w=\sin \theta_{\,\mathrm{i}}$. This allows to account for the non-uniform probability of randomly observing an AGN within a certain solid angle from the polar axis.

\section{The X-ray Baldwin effect}\label{sect:XBEslope}

Most of the studies on the X-ray Baldwin effect have found, using Eq.\,\ref{eq:fit_XBE}, a slope of $B\sim -0.2$ (e.g., \citealp{Iwasawa:1993ys}, \citealp{Page:2004kx}, \citealp{Bianchi:2007vn}). However, in all these works the values of $EW$ and $L_{\mathrm{X}}$ are obtained from individual observations of sources, and, as pointed out by \citet{Jiang:2006vn} and confirmed by \citet{Shu:2010zr,Shu:2012fk}, flux variability might play an important role in the observed X-ray Baldwin effect. \citet{Shu:2012fk} recently found a clear anti-correlation between the equivalent width of the iron K$\alpha$ line and the luminosity for individual sources which had several {\it Chandra}/HEG observations. Using a sample of 32 radio-quiet AGN, they also found that the fit {\it per source} (i.e. averaging different observations of the same object) results in a significantly flatter slope ($B=-0.11\pm0.03$) than that done {\it per observation} (i.e. using all the available observations for every source of the sample; $B=-0.18\pm0.03$).

In order to compare the slope obtained by the simulations in the two different luminosity bands with real data, we fitted the data per source (33 AGN) of \citeauthor{Shu:2010zr} (\citeyear{{Shu:2010zr}}; obtained fixing $\sigma=1\rm\,eV$) with Eq.\,\ref{eq:fit_XBE} in the Seyfert and quasar regime. Consistently to what reported in \citet{Shu:2010zr} we did not use the 3 AGN for which only upper limits of $EW$ were obtained. Using the 28 AGN in the Seyfert regime we obtained that the best fit to the X-ray Baldwin effect is given by $A=1.64\pm0.05$ and $B=-0.12\pm0.04$. {\it Chandra}/HEG observations are available for only 5 objects in the quasar regime, and at these luminosities we obtained $A=1.5\pm0.3$ and $B=-0.16\pm0.22$.

\begin{figure*}[t!]
\centering
\begin{minipage}[!b]{.48\textwidth}
\centering
\includegraphics[width=9cm]{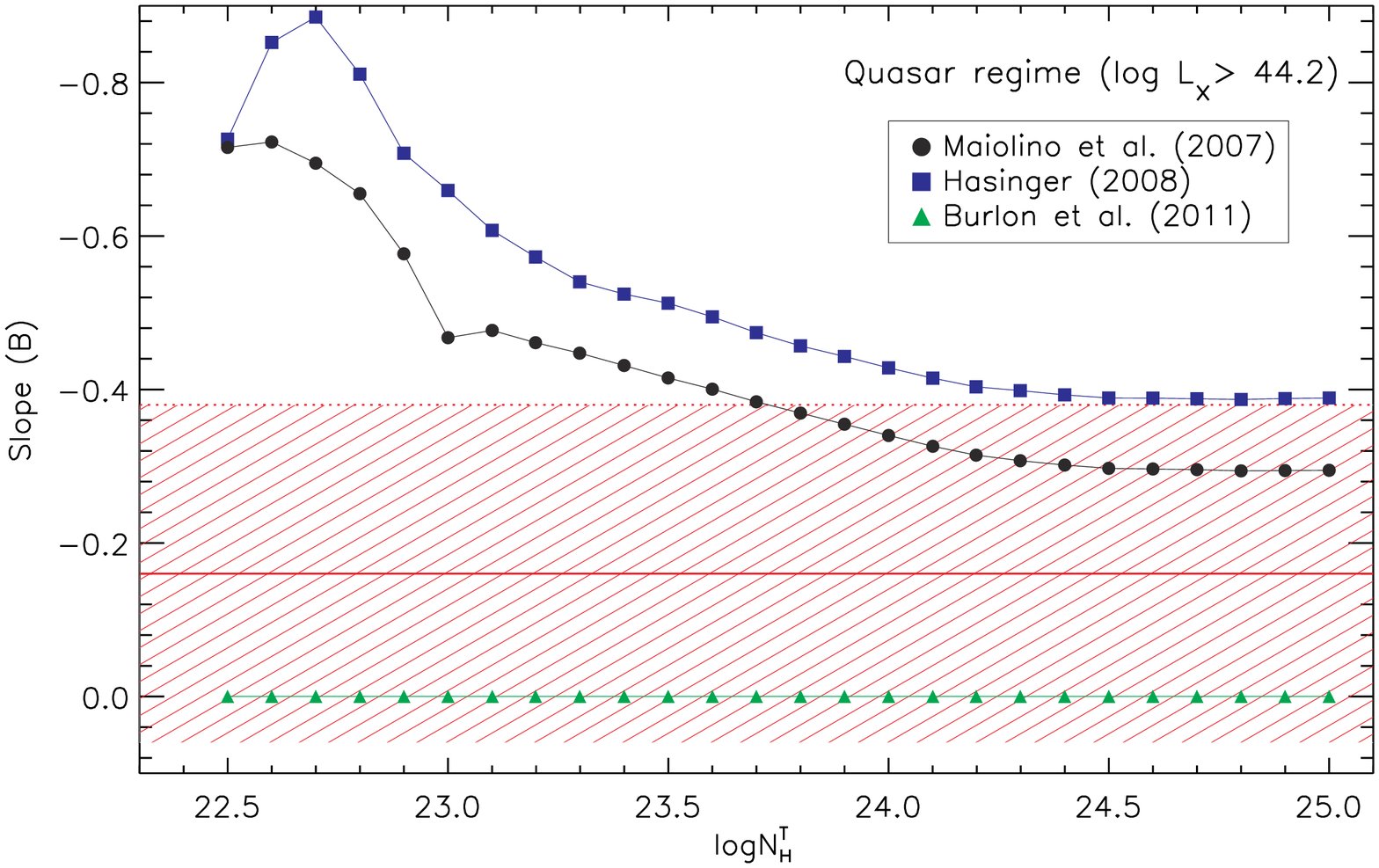}
\end{minipage}
\hspace{0.05cm}
\begin{minipage}[!b]{.48\textwidth}
\centering
\includegraphics[width=9cm]{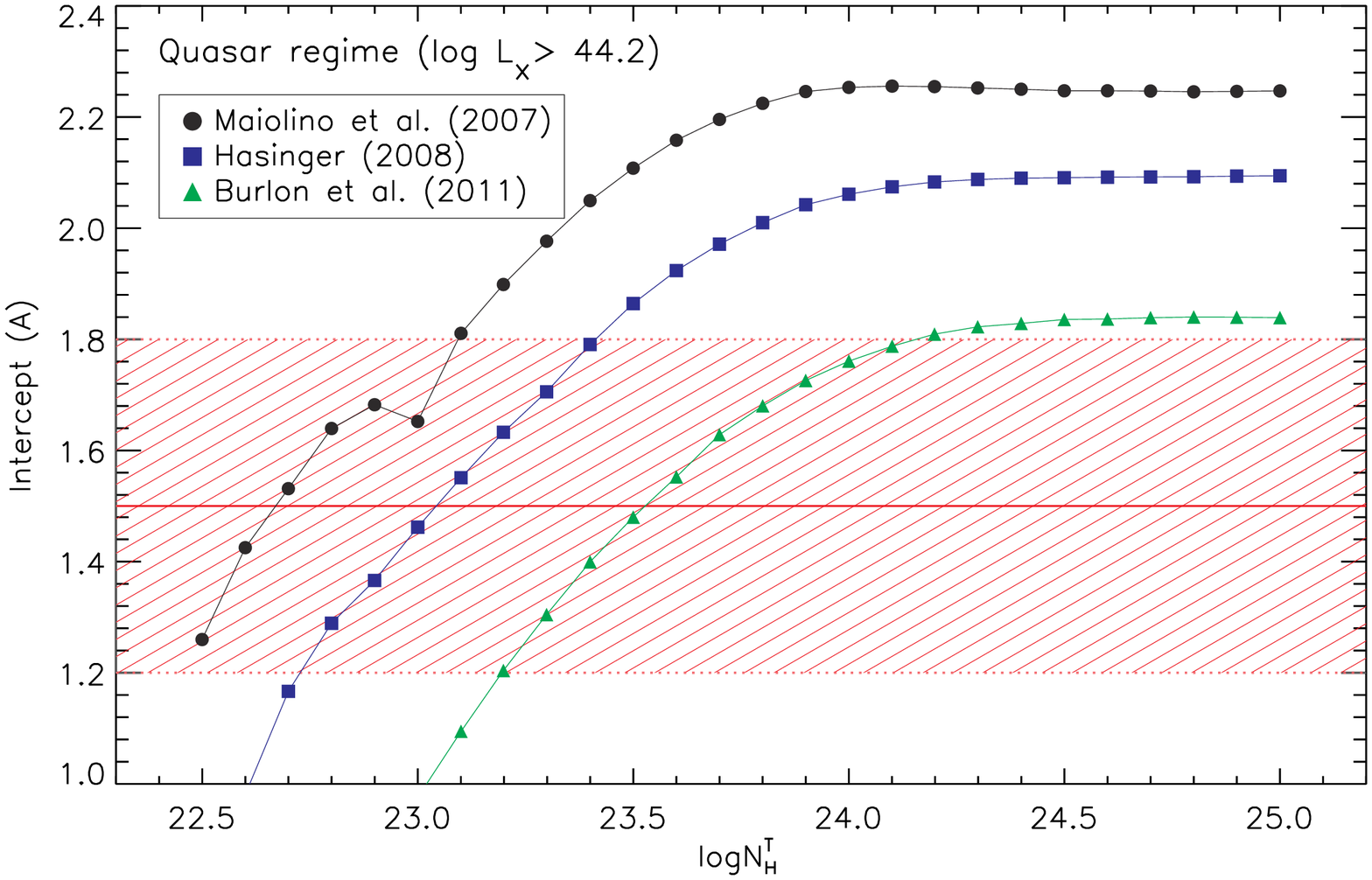}\end{minipage}

 \begin{minipage}[t]{1\textwidth}
  \caption{{\it Left panel}: value of the slope ($B$) of the X-ray Baldwin effect obtained simulating the variation of the reprocessed X-ray radiation with the luminosity for tori with different values of the equatorial column density $N_{\rm\,H}^{\rm\,T}$, for the different $\theta_{\mathrm{OA}}-L_{\mathrm{X}}$ relationships shown in Fig.\,\ref{fig:tor_CFL2} (Eq.\,\ref{Eq:Hasinger}-\ref{Eq:Maiolino}) in the quasar regime ($\log L_{\mathrm{X}} > 44.2$). The model used is that of \citet{Brightman:2011oq}, and the simulated data were fitted with Eq.\,\ref{eq:fit_XBE}. The red line represents the value of the slope ($B=-0.16\pm0.22$) obtained by fitting in the quasar regime the data {\it per source} reported in \citet{Shu:2010zr}; the red shadowed area represents the $1\sigma$ contour of the slope. {\it Right panel}: same as left panel, but considering the intercept ($A$) of the X-ray Baldwin effect obtained by our simulations. The red line and shadow represent the value of the intercept obtained by the fit to the {\it Chandra}/HEG data in the quasar regime and its $1\sigma$ error ($A=1.5\pm0.3$), respectively}
\label{fig:interc_sy_2}
 \end{minipage}

\end{figure*}

\subsection{Seyfert regime}\label{sec:Syregime}

Fitting the simulated data with Eq.\,\ref{eq:fit_XBE} in the Seyfert regime, we found that the slope obtained becomes flatter for increasing values of the equatorial column density of the torus (Fig.\,\ref{fig:slope_NH}, left panel). This is related to the fact that the iron K$\alpha$ $EW$ is tightly connected to both $N_{\rm\,H}^{\rm\,T}$ and  $\theta_{\mathrm{OA}}$, so that $EW$ is more strongly dependent on $\theta_{\mathrm{OA}}$ (and thus on its variation) for low values of $N_{\rm\,H}^{\rm\,T}$, which results in a steeper slope. On the other hand, when $N_{\rm\,H}^{\rm\,T}$ increases the dependence on $\theta_{\mathrm{OA}}$ becomes weaker, and the slope flatter. As shown in Fig.\,\ref{fig:slope_NH} (left panel), while the values of $B$ obtained by the $\theta_{\mathrm{OA}}-L_{\mathrm{X}}$ relationships of \citet{Burlon:2011cr} and \citet{Hasinger:2008ve} are similar along the whole range of $N_{\rm\,H}^{\rm\,T}$ considered, flatter slopes are obtained for that of \citet{Maiolino:2007bh}. In particular for the latter relationship the correlation becomes positive for $\log N_{\rm\,H}^{\rm\,T}\gtrsim 24$. 

By combining the observations of the X-ray Baldwin effect with our simulations we can extrapolate the average value of the equatorial column density of the torus of the unobscured AGN in the {\it Chandra}/HEG sample of \citet{Shu:2010zr}. A slope consistent within $1\sigma$ with our fit to the X-ray Baldwin effect in the Seyfert regime is obtained for $N_{\rm\,H}^{\rm\,T}\geq 10^{23.1}\rm\,cm^{-2}$ for the $\theta_{\mathrm{OA}}-L_{\mathrm{X}}$ relationship of \citet{Hasinger:2008ve}, for $N_{\rm\,H}^{\rm\,T}\gtrsim 10^{23.2}\rm\,cm^{-2}$ for that of \citet{Burlon:2011cr}, and for $N_{\rm\,H}^{\rm\,T}\lesssim10^{23}\rm\,cm^{-2}$ for the relationship of \citet{Maiolino:2007bh}. Comparing the value of the intercept obtained by the simulations with that resulting from our fit of the X-ray Baldwin effect in the Seyfert regime, we found that only a narrow range of average equatorial column densities of the torus can reproduce the observations (right panel of Fig.\,\ref{fig:slope_NH}). For the $\theta_{\mathrm{OA}}-L_{\mathrm{X}}$ relationship of \citet{Maiolino:2007bh} we found $22.9 \lesssim \log N_{\rm\,H}^{\rm\,T} \lesssim 23$, while for those of \citet{Burlon:2011cr} and \citet{Hasinger:2008ve} we obtained $\log N_{\rm\,H}^{\rm\,T} \simeq 23.2$. Using both the values of $A$ and $B$, it is possible to extract the average values of $N_{\rm\,H}^{\rm\,T}$ that can explain the X-ray Baldwin effect. As it can be seen from the two figures for both the $\theta_{\mathrm{OA}}-L_{\mathrm{X}}$ relationship of \citet{Hasinger:2008ve} and \citet{Burlon:2011cr} the only value of column density consistent with both the observed intercept and slope is $\log N_{\rm\,H}^{\rm\,T}\simeq 23.2$, while for the IR $\theta_{\mathrm{OA}}-L_{\mathrm{X}}$ relationship of \citet{Maiolino:2007bh} the values allowed are in the range $22.9 \lesssim \log N_{\rm\,H}^{\rm\,T} \lesssim 23$. A similar result is obtained studying the $A/B$ chi-squared contour plot of our fit to the {\it Chandra}/HEG data.  Using the model of \citet{Brightman:2011oq} to simulate the X-ray Baldwin effect in the Seyfert regime we obtained a range of $N_{\rm\,H}^{\rm\,T}$ consistent with that found using the model of \citet{Ikeda:2009nx} ($23.2 \lesssim \log N_{\rm\,H}^{\rm\,T} \lesssim 23.3$). The lower values of $N_{\rm\,H}^{\rm\,T}$ needed to explain the X-ray Baldwin effect using the relationship of \citet{Maiolino:2007bh} are due to the larger values of $f_{\mathrm{obs}}$ (and thus of $\theta_{\mathrm{OA}}$) predicted by Eq.\,\ref{Eq:Maiolino}. This is again due to the fact that $EW$ depends on both $N_{\rm\,H}^{\rm\,T}$ and  $\theta_{\mathrm{OA}}$, thus increasing the latter one would obtain lower values of the former.

\subsection{Quasar regime}\label{sec:Qsoregime}
To study the behavior of $B$ in the quasar regime, we fitted the simulated data using Eq.\,\ref{eq:fit_XBE} for each value of $N_{\rm\,H}^{\rm\,T}$ for luminosities $L_{\mathrm{X}}>L_{\mathrm{X}}^{\mathrm{Q}}$. Our simulations show that $EW$ decreases more steeply in the quasar regime than at lower luminosities for the $\theta_{\mathrm{OA}}-L_{\mathrm{X}}$ relationships of \citet{Hasinger:2008ve} and \citet{Maiolino:2007bh}, with values of the slope of $B\lesssim -0.3$ (left panel of Fig.\,\ref{fig:interc_sy_2}). The slopes obtained by the simulations are consistent with those found by fitting {\it Chandra}/HEG data in the quasar regime for the $\theta_{\mathrm{OA}}-L_{\mathrm{X}}$ relationship of \citet{Maiolino:2007bh} for $N_{\rm\,H}^{\rm\,T}\geq 10^{23.7}\rm\,cm^{-2}$. The values of the slope expected using the relationship of \citet{Hasinger:2008ve} are steeper than the observed value for the whole range of column densities considered, while the flattening of the relationship of \citet{Burlon:2011cr} at high luminosities results in a slope of $B\sim0$ along the whole range of $N_{\rm\,H}^{\rm\,T}$, consistent within 1$\sigma$ with the observations.  The intercepts obtained using the relationship of \citet{Maiolino:2007bh} are consistent with the observed value for $22.5 \leq \log N_{\rm\,H}^{\rm\,T} \leq 23.1$ (right panel of Fig.\,\ref{fig:interc_sy_2}), with no overlap with the values of $N_{\rm\,H}^{\rm\,T}$ needed by the slope. We obtained intercepts that are consistent with the observations for $ 22.7 \lesssim \log N_{\rm\,H}^{\rm\,T}\lesssim 23.4$ and $23.2 \lesssim \log N_{\rm\,H}^{\rm\,T}\lesssim 24.2$ for the relation of \citet{Hasinger:2008ve} and \citet{Burlon:2011cr}, respectively. Thus only the hard X-ray $\theta_{\mathrm{OA}}-L_{\mathrm{X}}$ relationship \citet{Burlon:2011cr} is able to explain, for average values of the equatorial column density of the torus  in the range $23.2 \lesssim \log N_{\rm\,H}^{\rm\,T}\lesssim 24.2$, at the same time both the intercept and the slope of the X-ray Baldwin effect at high luminosities.

\section{Discussion}\label{Sec:discussion}
Since its discovery about 20 years ago, the existence of the X-ray Baldwin effect has been confirmed by several works performed with the highest spectral resolution available at X-rays. So far, several possible explanations have been proposed. \citet{Jiang:2006vn} argued that the observed anti-correlation could be related to the delay of the reprocessed radiation with respect to the primary continuum. The response of the circumnuclear material to the irradiated flux is not simultaneous, and one should always take this effect into account when performing studies of reprocessed features as the iron K$\alpha$ line or the Compton hump. \citet{Shu:2010zr,Shu:2012fk} have shown that averaging the values of $L_{\mathrm{X}}$ and $EW$ for all the observations of each source results in a significantly flattened anti-correlation. However, by itself variability fails to fully account for the observed correlation \citep{Shu:2012fk}.  

\subsection{Explaining the X-ray Baldwin effect with a luminosity-dependent covering factor of the torus}
A mechanism often invoked to explain the X-ray Baldwin effect is the decrease of the covering factor of the molecular torus with the luminosity (e.g., \citealp{Page:2004kx}, \citealp{Bianchi:2007vn}). The decrease of the fraction of obscured sources with the luminosity has been reported by several works performed at different wavelengths in the last decade, although some discordant results have been presented (e.g., \citealp{Dwelly:2006fk}, \citealp{Lawrence:2010uq}). In this work we have showed that the covering factor-luminosity relationships obtained in the medium and hard X-ray band can explain well the X-ray Baldwin effect in the $10^{42}-10^{44.2}\rm\,erg\,s^{-1}$ luminosity range. In particular our simulations show that it is possible to reproduce the slope of the X-ray Baldwin effect with luminosity-dependent unification for average equatorial column densities of the torus of $\log N_{\rm\,H}^{\rm\,T}\gtrsim 23.1$, and both the slope and the intercept for  $\log N_{\rm\,H}^{\rm\,T}\simeq 23.2$ (Fig.\,\ref{fig:slope_NH}). In the same luminosity range the $\theta_{\mathrm{OA}}-L_{\mathrm{X}}$ IR relationship of \citet{Maiolino:2007bh} can explain the X-ray Baldwin effect for $22.9 \lesssim \log N_{\rm\,H}^{\rm\,T}\lesssim 23$.

In the quasar regime we have shown that, while the medium X-ray $\theta_{\mathrm{OA}}-L_{\mathrm{X}}$ relationship of \citet{Hasinger:2008ve} cannot explain the observations (Fig.\,\ref{fig:interc_sy_2}), the slope obtained by the IR $\theta_{\mathrm{OA}}-L_{\mathrm{X}}$ relationship is consistent with our fit to the {\it Chandra}/HEG data in the same luminosity band. However, as for the latter relationship the range of values of $N_{\rm\,H}^{\rm\,T}$ required to explain the slope and the intercept do not overlap, it cannot be considered as a likely explanation. The hard X-ray $\theta_{\mathrm{OA}}-L_{\mathrm{X}}$ relationship of \citet{Burlon:2011cr} is flat above $L_{\mathrm{X}}^{\mathrm{Q}}$, and would thus produce a constant $EW$ of the iron K$\alpha$ line and a slope of $B\sim0$ for the whole range of $N_{\rm\,H}^{\rm\,T}$, consistent within $1\sigma$ with the value found using {\it Chandra}/HEG data. The intercept we found using this relation is also consistent with the observational value for a large range of average equatorial column densities of the torus ($23.2 \lesssim \log N_{\rm\,H}^{\rm\,T}\lesssim 24.2$). The fraction of obscured AGN is not well constrained at high-luminosity, thus any trend with the luminosity between that of \citet{Burlon:2011cr} and that of \citet{Hasinger:2008ve} would be able to reproduce the observed slope. The relation of \citet{Hasinger:2008ve} produces negative values of $f_{\rm\,obs}$ for $\log L_{\mathrm{X}}\gtrsim 45.8$, thus a flattening of the decline is expected below this luminosity. From Fig.\,7 (left panel) in \citet{Hasinger:2008ve} one can see that indeed above $\log L_{\mathrm{X}}\simeq 45$ the value of $f_{\rm\,obs}$ appears to be constant, similarly to what has been found at hard X-rays. A flattening of $f_{\mathrm{obs}}$ in the quasar regime would also be expected when considering the large amount of accreting material needed to power the AGN at these luminosities. It must however be stressed that at high luminosities the X-ray Baldwin effect is not well studied, and the {\it Chandra}/HEG sample we used in this luminosity range is small, not allowing us to reach a firm conclusion on the variation of the iron K$\alpha$ $EW$ with the luminosity. Possible evidence of a flattening of the X-ray Baldwin effect in the quasar regime has been recently found by \citet{Krumpe:2010vn}. 

\subsection{On the differences between X-ray and IR half-opening angle-luminosity relationships}
The main difference between the different $\theta_{\mathrm{OA}}-L_{\mathrm{X}}$ relationships we used is that those obtained in the X-rays are based on direct observations of the absorbing material in the line of sight, while the one of \citet{Maiolino:2007bh} is extrapolated from the ratio of the thermal infrared emission to the primary AGN continuum. The $\theta_{\mathrm{OA}}-L_{\lambda}(5100\AA)$ relationship obtained by \citet{Maiolino:2007bh} was converted in $\theta_{\mathrm{OA}}-L_{\mathrm{X}}$ using the $L_{\,2\rm\,keV}-L{\lambda}(2500\AA)$ relation obtained by \citet{Steffen:2006uq}, and then converted in the $L_{\mathrm{X}}-L_{\lambda}(5100\AA)$ relation assuming the optical-UV spectral slope obtained by \citet{Vanden-Berk:2001kx}. All this is likely to introduce some error in the $\theta_{\mathrm{OA}}-L_{\mathrm{X}}$ obtained by their IR work. It has been argued by \citet{Maiolino:2007bh} that the larger normalization of the $\theta_{\mathrm{OA}}-L_{\mathrm{X}}$ relation they found is related to the fact that medium X-ray surveys such as that of \citet{Hasinger:2008ve} are likely to miss a certain fraction of heavily obscured objects, which can instead be detected in the IR. This is due to the fact that for Compton-thick AGN at energies $\lesssim 10\rm\,keV$ most of the X-ray emission is depleted. However, hard X-ray surveys as that of \citet{Burlon:2011cr}, which are much less biased by absorption, have found a normalization of the $\theta_{\mathrm{OA}}-L_{\mathrm{X}}$ relation consistent with that obtained at lower X-ray energies (see Fig.\,\ref{fig:tor_CFL1}). Our results also show that the X-ray Baldwin effect can be explained by a Compton-thin torus, and that larger values of $f_{\mathrm{obs}}$ would imply even lower average values of $N_{\rm\,H}^{\rm\,T}$, so that missing heavily obscured objects would not significantly affect our results. X-rays are also probably better suited to probe the material responsible for the iron K$\alpha$ line emission. The line can in fact be emitted by both gas and dust, and while the IR can probe only the latter, X-rays are able to infer the amount of both gas and dust. We thus conclude that for the purpose of our study the X-ray $\theta_{\mathrm{OA}}-L_{\mathrm{X}}$ relations are better suited than those extrapolated from IR observations.

It could be argued that the value of the hydrogen column density commonly used to determine the fraction of obscured objects (and thus as an indicator of the presence of the torus) is $N_{\rm\,H}\sim10^{22}\rm\,cm^{-2}$, while the torus is believed to have larger values of the equatorial column density, and that this obscuration might be related to the presence of dust lanes or molecular structures in the host galaxy (e.g., \citealp{Matt:2000kx}). However, \citet{Bianchi:2009fk}, using the results of \citet{Della-Ceca:2008fk}, have showed that a similar decrease of $f_{\mathrm{obs}}$ with the luminosity is obtained when setting this threshold to a larger value of line-of-sight column density. They found that the fraction of CT objects decreases with the luminosity as $f_{CT}\propto L^{-0.22}$, similarly to what is found for $f_{\mathrm{obs}}$ in the X-rays (e.g., \citealp{Hasinger:2008ve}, see Eq.\,\ref{Eq:Hasinger}). A similar result was obtained by \citet{Fiore:2009fk} from a {\it Chandra} and {\it Spitzer} study of AGN in the COSMOS field. This implies that the contribution of dust lanes or of galactic molecular structures to the observed $N_{\rm\,H}$ does not affect significantly the $f_{\mathrm{obs}}-L_{\mathrm{X}}$ relationships obtained in the X-rays.

\subsection{The equatorial column density of the torus}

\begin{figure}[t!]
\centering
\includegraphics[width=9cm]{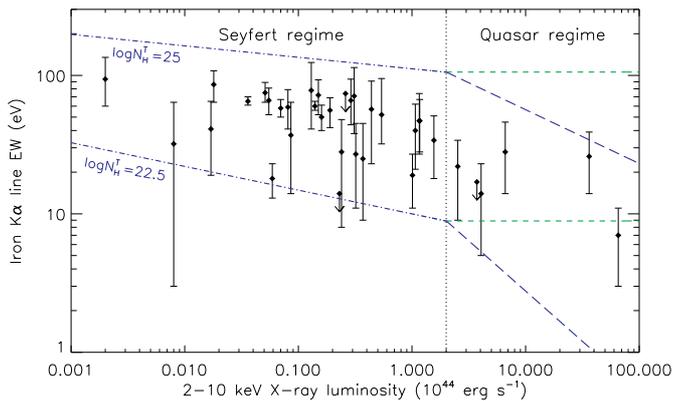}
\caption{Iron K$\alpha$ $EW$ versus X-ray luminosities and predicted trends obtained for different values of the equatorial column density of the torus. The points are the values of the equivalent width of the iron K$\alpha$ line reported by \citet{Shu:2010zr}, obtained by averaging multiple {\it Chandra}/HEG observations of AGN. The two blue dash-dotted lines are the fits to our simulations of the X-ray Baldwin effect using the $\theta_{\mathrm{OA}}-L_{\mathrm{X}}$ relationship of \citet{Hasinger:2008ve} for the Seyfert regime ($\log EW=1.01-0.17\log L_{\mathrm{X,44}} $ for $\log N_{\rm\,H}^{\rm\,T}=22.5$,  and  $\log EW=2.05-0.08\log L_{\mathrm{X,44}}$ for $\log N_{\rm\,H}^{\rm\,T}=25$). The blue long-dashed lines represent the $EW-L_{\mathrm{X}}$ relations obtained in the quasar regime ($\log EW=1.17 -0.73\log L_{\mathrm{X,44}} $ for $\log N_{\rm\,H}^{\rm\,T}=22.5$ and $\log EW=2.14-0.39\log L_{\mathrm{X,44}}$ for $\log N_{\rm\,H}^{\rm\,T}=25$) using the relationship of \citet{Hasinger:2008ve}, while the green dashed lines represent those obtained using the relationship of \citet{Burlon:2011cr}. The intercepts obtained in the quasar regime have been modified in order to match those obtained at lower luminosities.}
\label{fig:Chandra_model}
\end{figure}%

The distribution of values of the equatorial column density of the torus in AGN is still poorly constrained. X-ray observation can in fact infer solely the obscuration in the line of sight, and only studies of the reprocessed X-ray emission performed using physical torus models like those of \citet{Ikeda:2009nx}, \citet{Brightman:2011oq} and \citet{Murphy:2009uq} can help to deduce the value of the equatorial column density of the torus. However, this kind of studies are still very scarce (e.g., \citealp{Rivers:2011fk,Brightman:2012fk}), besides being largely geometry-dependent. Studies of AGN in the mid-IR band performed using the clumpy torus formalism of \citet{Nenkova:2008kx} have shown that the number of clouds along the equator is $N_0\sim 5-10$ \citep{Mor:2009fk}. If as reported by \citet{Mor:2009fk} each cloud of the torus has an optical depth of $\tau_V\sim 30-100$ (i.e., $\log N_{\rm\,H}^{\rm\,T}\sim 22-23$), the equatorial column density of the torus is expected to be $\log N_{\rm\,H}^{\rm\,T}\sim 22.5-24$, in agreement with our results. It must however be noticed that the value of the intercept of the X-ray Baldwin effect obtained by the simulations is strongly dependent on our assumptions on the metallicity of the torus. This implies that our constraints of the average value of $N_{\rm\,H}^{\rm\,T}$ are also tightly related to the choice of the metallicity: lower values of the metallicity would lead to larger values of $N_{\rm\,H}^{\rm\,T}$. To study this effect we repeated our study in the Seyfert regime using half-solar metallicities for the reflection model. We obtained that in this scenario, in order to explain the X-ray Baldwin effect, one needs values of the equatorial column density of the torus about two times larger than for the solar-metallicity case ($23.5\lesssim \log N_{\rm\,H}^{\rm\,T} \lesssim 23.7$). To have a Compton-thick torus one would then need values of the metallicity of $Z\lesssim 0.2\,Z_{\,\sun}$. A Compton-thick torus has often been invoked to explain the Compton hump observed in the spectrum of many unobscured AGN (e.g., \citealp{Bianchi:2004fk}). It is still unclear however which fraction of the Compton hump is produced in the distant reflector and which in the accretion flow. From our study we have found that in unobscured objects the X-ray Baldwin effect can be explained by a luminosity-dependent covering factor of the torus for an average value of the equatorial column density of $\log N_{\rm\,H}^{\rm\,T}\simeq 23.2$. This value is lower than the line-of-sight $N_{\rm\,H}$ of many Seyfert\,2s (e.g., \citealp{Ricci:2011zr,Burlon:2011cr}), and it might be related either to the geometry we adopted or to presence of objects with sub-solar metallicities. In particular, due to the constant reflection angle relative to a local normal in any point of the reflecting surface, the spherical-toroidal geometry produces larger values of $EW$ (and thus larger values of the intercept) with respect to a toroidal structure \citep{Murphy:2009uq}.

It is possible that there exists a wide spread of equatorial column densities of the torus, thus one could envisage that this, together with the different values of $\theta_{\,\mathrm{i}}$, would introduce the scatter observed in the anti-correlation. In Fig.\,\ref{fig:Chandra_model} we show the X-ray Baldwin effect obtained from our simulations using the relationship of \citet{Hasinger:2008ve} for $\log N_{\rm\,H}^{\rm\,T}=22.5$ and $\log N_{\rm\,H}^{\rm\,T}=25$, in both the Seyfert and the quasar regime, together with the time-averaged {\it Chandra}/HEG data of \citet{Shu:2010zr}. From the figure it is evident that all the data are well within the range expected from our simulations, both in the Seyfert and in the quasar regime. It is important to remark that we do not know whether there exists a relation between the equatorial column density of the torus and the AGN luminosity. However, we have shown that the variation of the covering factor of the torus with the luminosity alone can fully explain the observed trend, so that no additional luminosity-dependent physical parameter is needed.

\subsection{Luminosity-dependent unification of AGN}
The relation of $f_{\mathrm{obs}}$ with the luminosity might be connected to the increase of the inner radius of the torus with the luminosity due to dust sublimation. Both near-IR reverberation \citep{Suganuma:2006fk} and mid-IR interferometric \citep{Tristram:2011uq} studies have confirmed that the inner radius of the molecular torus increases with the luminosity as $R_{\mathrm{in}}\propto L^{0.5}$. Considering the geometry of Fig.\,\ref{fig:tor_CFL1}, the fraction of obscured objects (see Eq.\,\ref{Eq:fractioncos}) would be related to ratio of the height to the inner radius of the torus ($H/R$) by
\begin{equation}
f_{\mathrm{obs}}\simeq\frac{H}{R}\sqrt{\frac{1}{1+(H/R)^2}}.
\end{equation}
For values of $H/R\lesssim 1$, $f_{\mathrm{obs}}\propto H/R$. In the original formulation of the {\it receding torus} model \citep{Lawrence:1991vn}, the height $H$ was considered to be constant. Assuming that $R$ has the same luminosity-dependence as $R_{\mathrm{in}}$, this would lead to $f_{\mathrm{obs}}\propto L^{-0.5}$. This has been shown to be inconsistent with the recent observations, which point towards a flatter slope ($f_{\mathrm{obs}}\propto L^{-0.25}$, e.g., \citealp{Hasinger:2008ve}, see Eq.\,\ref{Eq:Hasinger}), and which would imply $H\propto L^{0.25}$. \citet{Honig:2007kx} have shown that in the frame of {\it radiation-limited clumpy dust torus} model, one would obtain $H\propto L^{0.25}$, in agreement with the observations. 

The decrease of the covering factor with luminosity might also have important implications on the AGN dichotomy. It has been shown that Seyfert 2s appear to have on average lower luminosities and lower Eddington ratios (e.g., \citealp{Beckmann:2009fk}, \citealp{Ricci:2011zr}) than Seyfert\,1s and Seyfert\,1.5s, which suggests that they have on average a torus with a larger covering factor. This idea is also supported by the fact that the luminosity function of type-II AGN has been found to peak at lower luminosities than that of type-I AGN (\citealp{Della-Ceca:2008fk}, \citealp{Burlon:2011cr}), and by the results obtained by the recent mid-IR work of \citeauthor{Ramos-Almeida:2011fk} (\citeyear{Ramos-Almeida:2011fk}, see also \citealp{Elitzur:2012vn}). A difference in the torus covering factor distribution between different types of AGN would also explain why the average hard X-ray spectrum of Compton-thin Seyfert\,2s shows a larger reflection component than that of Seyfert\,1s and Seyfert\,1.5s \citep{Ricci:2011zr}. A similar result was found by \citet{Brightman:2012fk}: studying high-redshift AGN in the {\it Chandra} Deep Field South they found that more obscured objects appear to have tori with larger covering factors, although they did not find a clear luminosity dependence.

\section{Summary and conclusions}\label{Sec:summary}
In this work we have studied the hypothesis that the X-ray Baldwin effect is related to the decrease of the covering factor of the torus with the luminosity. We have used the physical torus models of \citet{Ikeda:2009nx} and \citet{Brightman:2011oq} to account for the reprocessed X-ray radiation, and the values of the fraction of obscured sources obtained by recent surveys in the X-rays and in the IR as a proxy of the covering factor of the torus. Our simulations show that the variation of the covering factor of the torus with the luminosity can explain the X-ray Baldwin effect. In the Seyfert regime ($L_{\mathrm{X}}\leq 10^{44.2}\rm\,erg\,s^{-1}$), the observed $EW-L_{\mathrm{X}}$ trend can be exactly (both in slope and intercept) reproduced by an average value of the equatorial column density of the torus of $\log N_{\rm\,H}^{\rm\,T}\simeq 23.2$, while a slope consistent with the observations is obtained for a larger range of column densities ($\log N_{\rm\,H}^{\rm\,T}\gtrsim 23.1$). At higher luminosities ($L_{\mathrm{X}}> 10^{44.2}\rm\,erg\,s^{-1}$) the situation is less clear due to the small number of high-quality observations available. Moreover, it is not clear whether in the quasar regime $f_{\mathrm{obs}}$ still decreases with the luminosity similarly to what is found in the Seyfert regime (as shown by \citealp{Hasinger:2008ve}), or is constant (as found by \citealp{Burlon:2011cr}). A flattening of the $\theta_{\mathrm{OA}}-L_{\mathrm{X}}$ relationship would be able to explain the observations at high luminosities (for $23.2 \lesssim \log N_{\rm\,H}^{\rm\,T}\lesssim 24.2$), and it might naturally arise from the large amount of accreting mass needed to power these luminous quasars.

In the next years {\it ASTRO-H} \citep{Takahashi:2010uq}, with its high energy-resolution calorimeter SXS, will allow to constrain the origin of the narrow component of the iron K$\alpha$ line, being able to separate even better than {\it Chandra}/HEG the narrow core coming from the torus from the flux emitted closer to the central engine. {\it ASTRO-H} will also be able to probe the narrow iron K$\alpha$ line in the quasar regime, allowing to understand the behavior of the X-ray Baldwin effect at high luminosities.

\begin{acknowledgements}
We thank the anonymous referee for his/her comments that helped improving the paper.
We thank XinWen Shu and Rivay Mor for providing us useful details about their work, Chin Shin Chang and Poshak Gandhi for their comments on the manuscript. CR thanks the Sherpa group and IPAG for hospitality during his stay in Grenoble. CR is a Fellow of the Japan Society for the Promotion of Science (JSPS).
 \end{acknowledgements}
 
 \bibliographystyle{aa}
 \bibliography{BaldwinI}
 
 \end{document}